\documentclass[aip,amsmath,amssymb,reprint]{revtex4-1}

\usepackage{graphicx}
\usepackage{dcolumn}
\usepackage{bm}
\usepackage[utf8]{inputenc}
\usepackage[T1]{fontenc}
\usepackage{mathptmx}
\usepackage{etoolbox}
\usepackage{float}
\usepackage{siunitx}
\usepackage{ulem}

\makeatletter
\def\@email#1#2{%
 \endgroup
 \patchcmd{\titleblock@produce}
  {\frontmatter@RRAPformat}
  {\frontmatter@RRAPformat{\produce@RRAP{*#1\href{mailto:#2}{#2}}}\frontmatter@RRAPformat}
  {}{}
}%
\makeatother
\begin{document}

\title{Fast-response low power atomic oven for integration into an ion microchip}
\author{Vijay~Kumar}
\affiliation{These authors contributed equally to this work.}
\affiliation{Sussex Centre for Quantum Technologies, University of Sussex, Brighton, BN1 9RH, U.K.}
\author{Martin~Siegele-Brown}
\affiliation{These authors contributed equally to this work.}
\affiliation{Sussex Centre for Quantum Technologies, University of Sussex, Brighton, BN1 9RH, U.K.}
\author{Parsa~Rahimi}
\affiliation{Sussex Centre for Quantum Technologies, University of Sussex, Brighton, BN1 9RH, U.K.}
\author{Matthew~Aylett}
\affiliation{Sussex Centre for Quantum Technologies, University of Sussex, Brighton, BN1 9RH, U.K.}
\author{Sebastian~Weidt}
\affiliation{Sussex Centre for Quantum Technologies, University of Sussex, Brighton, BN1 9RH, U.K.}
\affiliation{Universal Quantum Ltd, Brighton, BN1 6SB, U.K.}
\author{Winfried~Karl~Hensinger}
\affiliation{Sussex Centre for Quantum Technologies, University of Sussex, Brighton, BN1 9RH, U.K.}
\affiliation{Universal Quantum Ltd, Brighton, BN1 6SB, U.K.}
\email{w.k.hensinger@sussex.ac.uk}

\date{\today}

\begin{abstract}
We present a novel microfabricated neutral atom source for quantum technologies that can be easily integrated onto microchip devices using well-established MEMS fabrication techniques, and contrast this to conventional off-chip ion loading mechanisms. The heating filament of the device is shown to be as small as $90\times$\SI{90}{\micro\meter\squared}. Testing of the $^{171}$Yb fluorescence response is found to be in the low tens of milliseconds, two orders of magnitude faster compared to previous literature at a power of milliwatts making it desirable for low-power device packages. We demonstrate how the evaporation material can be capped in vacuum to work with materials such as Ba that oxidise easily in air, which can avoid the need for ablation lasers in the loading process. We calculate oven lifetimes to be over 10 years of continuous use for commonly used ion species in quantum technology. 
\end{abstract}

\maketitle

The advent of quantum technology has demonstrated many use cases for individual trapped ions and trapped atoms in small scale devices. Some of the areas that can best utilise these include, but are not limited to, quantum computing\cite{blueprint}, sensing\cite{sensing}, and metrology\cite{clock}. However, to best realise a quantum device using  trapped ions and atoms, scalability and deployability must be taken into account. This includes careful considerations as to how the device is designed, kept under vacuum, maintained, and loaded. Furthermore, the ion/atom loading mechanism should provide a high trapping rate without adding complexity and size to the device. 

In typical ion trapping experiments, atomic ions are generated by photoionisation of thermally sublimated neutral atoms \cite{gulde2001simple} or magneto-optically trapped (MOT) atoms \cite{sage2012loading}, or by direct loading of ionised atoms assisted by laser ablation \cite{leibrandt2007laser}. This ablation approach is more common for materials that oxidise easily in air, as a metal salt can be used as an ablation target. In many experimental configurations, the neutral atom source is positioned to the side of the trap surface. In other setups, backside loading - with the atom sources beneath the substrate - is used to prevent electrical shorts between small electrode gaps due to contaminants, and to scale the trapping system \cite{amini2010toward}. Efficient loading of ions from atomic ovens has been demonstrated \cite{gao2021optically} predicting that \SI{1}{\micro\gram} of calcium would allow for a decade of continuous operation.

\begin{figure}[b!]
\centering
\includegraphics[width=0.45\textwidth]{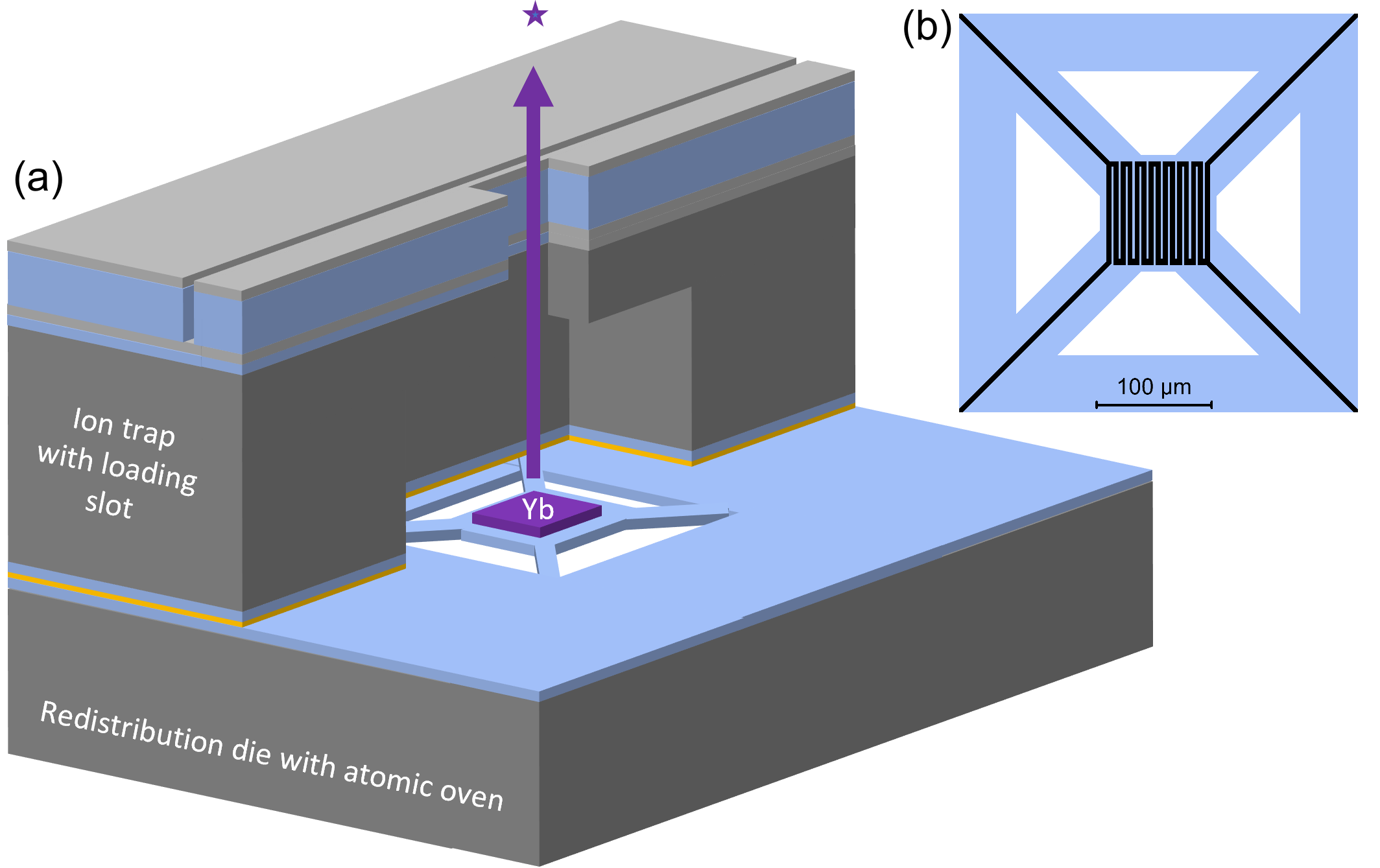}
\caption{(a) Schematic showing an ion trap with an integrated atomic oven for loading ytterbium. (b) Design of the atomic oven prototype.}
\label{fig:oven_schematic}
\end{figure}

Low-power, fully CMOS-compatible micro-hotplates have already been demonstrated for gas sensor applications \cite{Siegele}. Offering power consumption as low as \SI{20}{mW} at \SI{400}{\degreeCelsius}, the power dissipation of these micro-heaters is negligible compared to the rest of the ion trap. CMOS-compatible micro-hotplates will be useful for a scalable quantum computer, and can be integrated into a wafer that houses other components such as DACs or photodetectors. However, for a prototype, a simple platinum-based micro-hotplate is much more suitable, due to less overhead for the CMOS process and more flexibility of layer thickness. Loading from a \SI{1.5}{mm}-diameter microfabricated hotplate-based silicon-based atomic oven\cite{manginell2012situ, schwindt2016highly} and an even larger fused silica-based atomic oven\cite{pick2024low} located externally have been demonstrated, but neither oven is suitable for integration into a microfabricated ion trap.

In order to facilitate greater scalability, here we demonstrate the use of micro-scale atomic ovens as an atom source, microfabricated using well-known
MEMS techniques, and propose a use case of how this can be integrated into an ion-trap microchip by die or wafer bonding as shown in Figure \ref{fig:oven_schematic}. The microfabrication process of this atomic oven consists of fabricating plate-shaped micro-heaters and depositing ytterbium source material, which generates an atomic flux when heated on the heater plate. In this approach the microfabricated atomic ovens can be easily wafer or die bonded to a ion trap microchip with a backside loading slot to enable the atomic flux originating from the back of the substrate into the loading zone. By integrating this on-chip we significantly reduce the footprint compared to conventional ion loading mechanisms and also provide orders of magnitude faster response times compared to previous thermally sublimated atom sources for which a short response time of approximately \SI{12}{s}\cite{ox} is described for a conventional macroscopic oven controlled with a digital feedback loop implemented in a microcontroller. Furthermore, we show that capping the oven in vacuum did not affect the measured atomic flux, showing this oven can be used with materials that oxidise easily in air such as Ba, providing a significantly smaller overall device package compared to laser ablation loading by avoiding the need for an ablation laser. 

\section{Design}


The atomic flux of an oven required to achieve a certain number density $N(T)$, is given approximately as \cite{gao2021optically}
$
J_{atom}(T)=
N(T)\bar{v}(T) \pi r^2_0,
$
where $T$ is the temperature of the oven, $\bar{v}$ the mean thermal velocity of the atoms, and $r_0$ the distance to the oven. 
With an integrated oven having a distance of approximately \SI{0.5}{mm} to the ion trap, compared to tens of millimeters for an external oven, long lifetimes can be achieved with relatively small sources. An atom flux $J_{atom} = \num{38e3}$ results in a number density of \SI{100}{cm^{-3}} which is sufficient to trap a Ca ion within seconds \cite{gao2021optically}. Trapping rates are proportional to the number density \cite{Wunderlich2007}.
The expected flux Q from a surface is calculated in g$\cdot$cm$^{-2}$$\cdot$s$^{-1}$ using the equation\cite{evaprate} $Q=0.058P_v(T)\sqrt{M/T}$ where M is the molecular weight, T is the temperature in K and $P_v(T)$ is the vapour pressure of the metal at a given temperature in Torr. Alcock et al. \cite{Alcock} give the vapour pressure as $\text{log}_{10}(p/\SI{}{atm}) = A - BT^{-1}- C\text{log}T$ reproducing the vapour pressure with an accuracy of \SI{5}{\%} or better. The required temperature to achieve a number density in the trapping location of 100, $10^5$, and \SI{e8}{\per\cubic\centi\meter} is given in Table \ref{tab:my_label} for a selection of elements.

\begin{table}[b!]
    \centering
\begin{tabular}{lccccccc}
\hline
Element & Ba & Be & Ca & Cd & Mg & Sr & Yb \\
N [\SI{}{\per\cubic\centi\meter}] & \multicolumn{7}{c}{Temperature T [K]} \\
\hline
100 & 461 & 802 & 451 & 278 & 369 & 411 & 385 \\
$10^5$ & 545 & 940 & 530 & 326 & 433 & 484 & 452 \\
$10^8$ & 669 & 1135 & 646 & 394 & 525 & 591 & 549 \\
\hline
\end{tabular}
    \caption{Temperature [K] to achieve a number density in the trapping location of 100, $10^5$, and \SI{e8}{\per\cubic\centi\meter} for different elements commonly used in ion trapping.}
    \label{tab:my_label}
\end{table}

Depending on the element, 0.5 to \SI{1.5}{ng} are required for continuous operation over ten years for a number density of \SI{100}{\per\cubic\centi\meter}.
A low-duty cycle operation would increase the lifetime and the micro-heater can heat up within tens of milliseconds. However, only source material in sight of the ion can be considered active. For a loading slot aperture at \SI{175}{\micro\meter} from the ion trapping position, and a distance of \SI{330}{\micro\meter} from the oven to the aperture, the active area\footnote{Area in line of sight of the trapping position} of the oven  is $1+330/175\approx 3$ times the size of the loading slot. However, even for a small loading slot of \SI{15}{\micro\meter} x \SI{5}{\micro\meter}, a layer thickness of only 182 to \SI{780}{nm} is required. Although making the micro-hotplates larger than the active area does not improve oven lifetimes, larger micro-hotplates allow easier deposition of the source material, and allow lower current densities for the heating element.


The resistivity and maximum current density of the heating element limit the maximum power. By substituting the current density $I=j w t$ and resistivity $R=\rho l /(w t)$ into Ohm's law it can be shown easily that the power dissipation $P=I^2R=(j w t)^2\rho l/(w t)=j^2\rho w l t=j^2\rho A t$ is a function of the current density $\rho$, area $A=wl$, and thickness $t$ of the heating element, and is design-independent. Electromigration is an important issue for micro-heaters, as these are operated at high temperatures. The literature\cite{courbat2008reliability} gives a maximum current density of \SI{\approx3.5e5}{A/cm^2} for platinum-based heating elements. Using a \SI{500}{nm}-thick platinum layer, we need an area of \SI{\approx1500}{\micro\meter\squared} for \SI{1}{mW} heating power.
There has been extensive research into heating element designs capable of providing good temperature uniformity \cite{Siegele, lee2003design}. These are designed in such a way as to compensate for the fact that most of the thermal flux is located at the corners, due to heat conduction through the arms and much lower heat flux due to thermal radiation and convection through air.

\section{Fabrication of atomic oven}

\begin{figure}[b!]
\centering

         \includegraphics[width=0.49\textwidth]{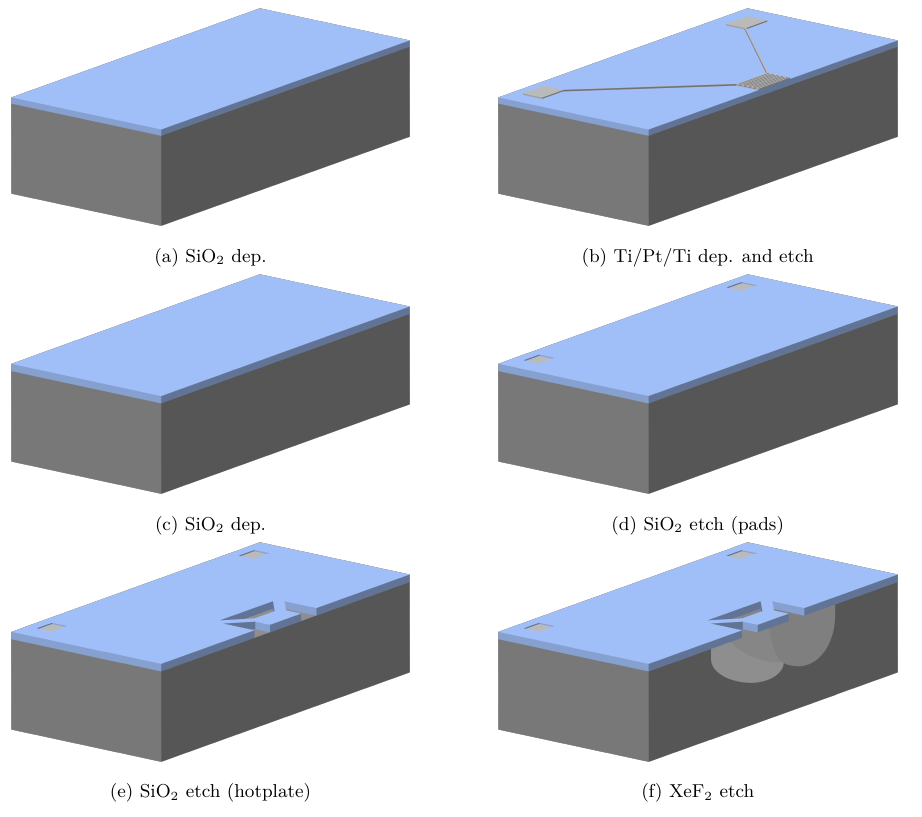}

\caption[Fabrication process flow for micro-heaters.]{Fabrication process flow for micro-heaters. Note that the adhesion layers and photolithography steps are not shown in the schematics. Positive photoresist was used for all the lithography steps. All schematics not to scale.}
\label{fig:hotplate_pf}
\end{figure}

The fabrication of the atomic oven is described in this section. The main structural part of the the micro-hotplate is formed by the \SI{2}{\micro\meter}-thick wet oxide that is thermally grown on the wafer (Fig.~ \ref{fig:hotplate_pf} (a)). On top, a 10/500/\SI{10}{nm}-thick Ti/Pt/Ti layer is sputtered and photographically patterned using ion beam etching (Fig.~\ref{fig:hotplate_pf} (b)). After resist stripping, a \SI{1}{\micro\meter}-thick PECVD silicon oxide layer is deposited as passivation (Fig.~\ref{fig:hotplate_pf} (c)). To integrate easily with ion trap microchips we can deposit an Au layer at this point and etch this using ion beam etching to expose the hotplate and bonding pads. This allows for easy die bonding onto our ion trap chip with a loading slot. One dry etching step is used to etch down to the bonding pads of the heater through the silicon dioxide (Fig.~\ref{fig:hotplate_pf} (d)), and another to define the structure of the micro-hotplate (Fig.~\ref{fig:hotplate_pf} (e)). These steps can be combined, but depending on the thickness of the wet oxide there can be a long exposure of the Ti/Pt/Ti layer to the etching chemistry, which can cause problems such as resputtering or contamination of the chamber. After resist removal, the hotplate structure is released using XeF$_2$ etching (Fig.~\ref{fig:hotplate_pf} (f)). As XeF$_2$ is highly selective between Si and all other materials used, no photoresist is necessary. This leaves the heating element floating and thermally isolated from the Si substrate. This step completes fabrication of the micro-hotplate and the ytterbium can be deposited along with any capping layers. 
We sputtered \SI{500}{nm} Yb onto the oven utilising a shadow mask (\textit{LESKER} PVD 75 \& Yb sputter target EJTYBXX303A2). On some ovens we additionally deposited \SI{\approx5}{nm} Bi or ITO on top of the Yb without breaking vacuum, as a capping layer to protect the Yb against oxidation. This becomes more important when using elements such as Ba that rapidly oxidise in air. Bi and ITO were chosen as capping layers due to their relatively low evaporation temperatures at a pressure of \SI{e-6}{mbar} of approximately \SI{410}{\degreeCelsius} and \SI{500}{\degreeCelsius} respectively.

\begin{figure}[t!]
\centering
         \centering
         \includegraphics[width=0.35\textwidth]{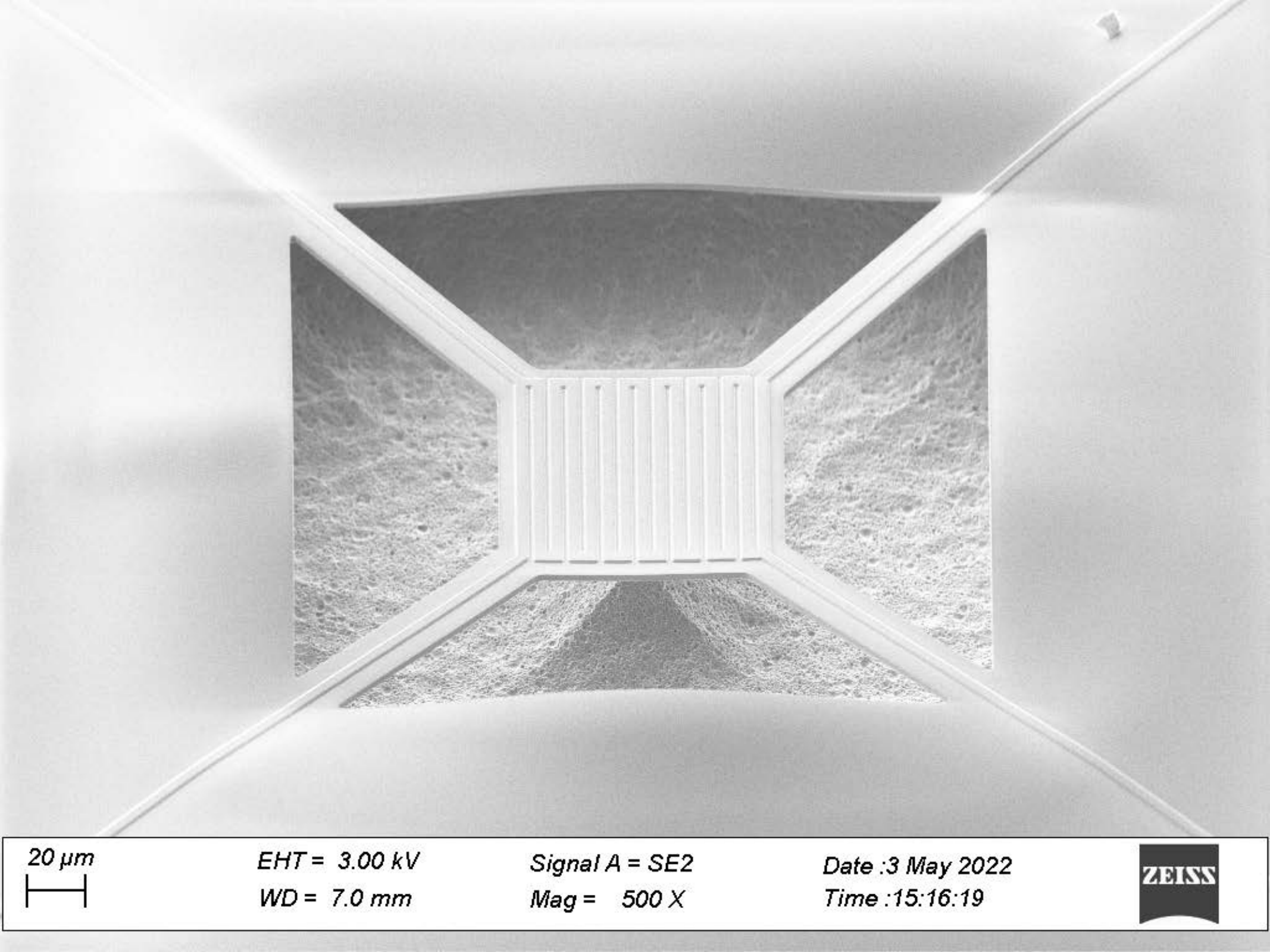}

\caption{SEM images with \SI{500}{x} magnification of the fabricated micro-hotplate before depositing ytterbium.}
\label{fig:ovenSEM}
\end{figure}

\begin{figure}[b!]
\centering
\includegraphics[width=0.45\textwidth]{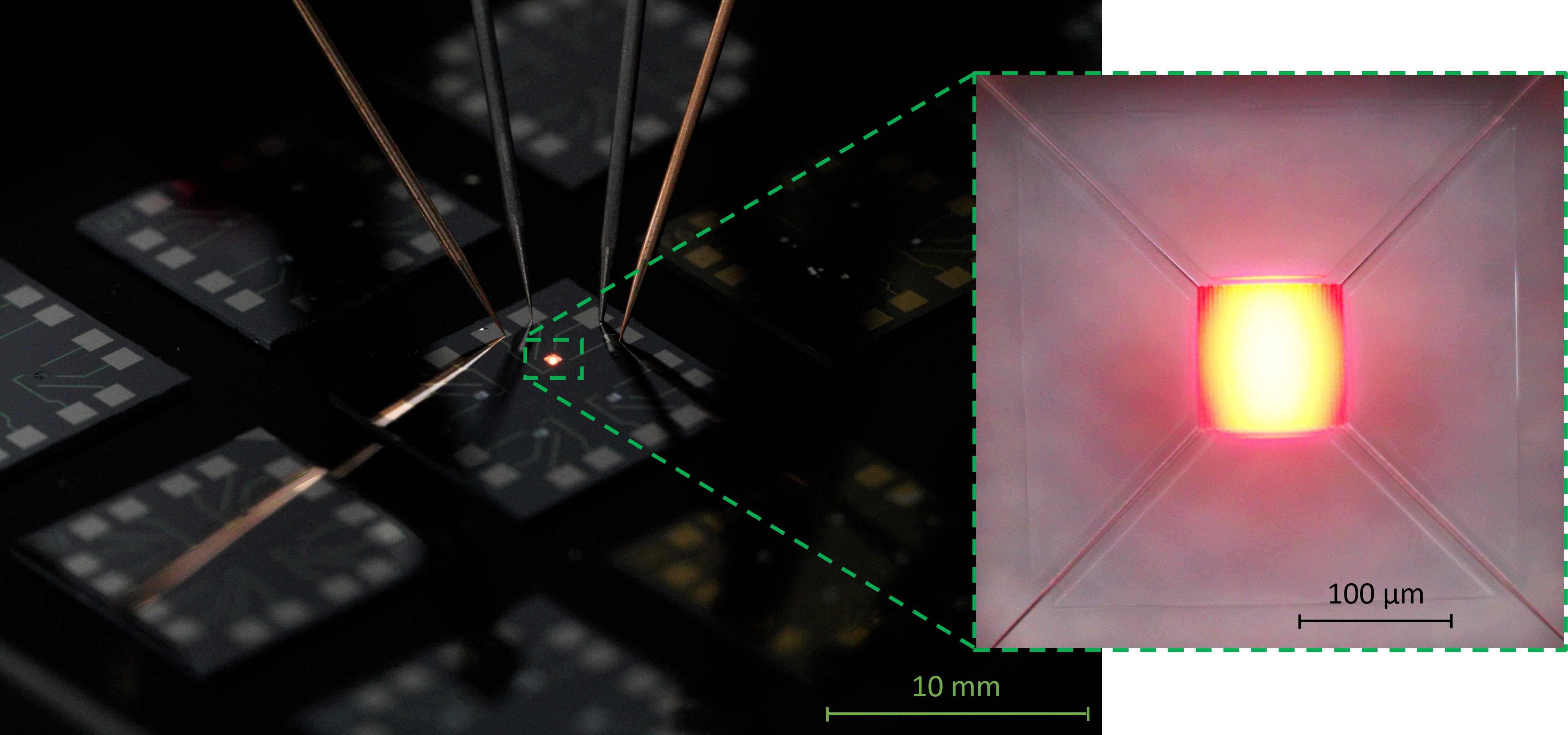}
\caption{Photograph of heated micro-hotplate with inlay showing a 20x magnification microscope image of the hotplate at the power dissipation of \SI{38}{mA} operated in air.}
\label{fig:uhotplate_pic}
\end{figure}

Figure \ref{fig:ovenSEM} shows an SEM image of the fabricated micro-hotplate at \SI{500}{x} magnification. Figure \ref{fig:uhotplate_pic} shows a photograph of a heated micro-hotplate with the inlay showing a \SI{20}{x} magnification microscope image of the hotplate at the power dissipation of \SI{38}{mW} operated in air, where the main mode of heat transport is through air. Thermal glow was first observed at \SI{24}{mW} power dissipation. 

\section{Testing}
A fluorescence test was carried out to verify that atomic flux was emitted from the micro oven. The test system consists of a hexagonal vacuum system with a PCB on which a micro oven test chip is mounted. The oven is connected to the PCB traces with ribbon bonds that are wired to a vacuum feed-through with copper wire. One laser port allows laser access perpendicular to the oven surface, and another on the top of the chamber allows imaging of the oven from the top down. A \SI{399}{nm} wavelength laser beam is collimated out of a fibre. The laser beam is aligned to pass over the micro oven position. An ion gauge in vacuum allows for accurate monitoring of the pressure of the chamber. The test was carried out at a pressure of \SI{9e-7}{mbar}. The oven is connected to a variable DC power supply and ammeter in series.  

\begin{figure}[b!]
\centering

         \includegraphics[width=0.5\textwidth]{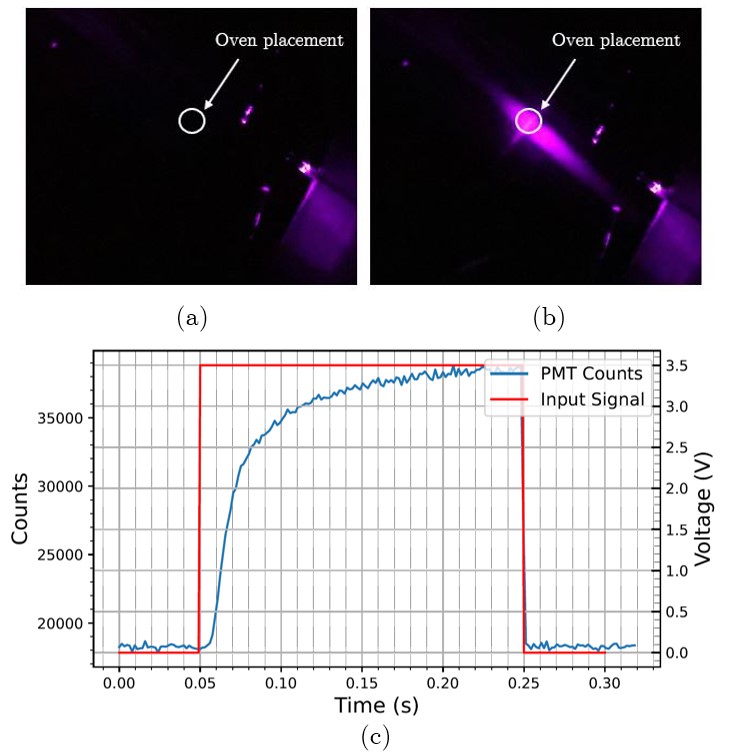}
         \caption{Micro oven fluorescence test with the oven turned off (a) and with 
         current applied showing strong fluorescence (b). (c) shows the fluorescence response of a microfabricated oven. The red line denotes the input voltage and blue shows the counts received at the PMT. This data is averaged over 200 runs.
}
         \label{fig:response}
\end{figure}

It is well-known that the Yb $^1$S$_0$-$^1$P$_1$ transition is excited by a \SI{399}{nm} wavelength and also fluoresces at this same wavelength \cite{399flo}. Therefore, we expect to see an increase in \SI{399}{nm} light intensity when the atomic flux from the micro oven hits the laser beam. To image this a conventional CMOS camera is used that is modified by removing the built-in UV filter. The results of this are shown in Figure \ref{fig:response}.

To compare the fluorescence response time to the literature, an experiment was set up to trigger a voltage pulse accurately and detect fluorescence on an appropriate time scale. The test was carried out in the hexagonal vacuum chamber as described before, with the microfabricated oven mounted on a PCB in the same way inside the vacuum chamber. The optimal voltage was found to achieve weak fluorescence by applying a voltage using a variable power supply. A PMT (photomultiplier tube) was positioned into the oven. Using an FPGA we can trigger a \SI{3.5}{V} pulse from a signal generator and record the PMT counts over this time period. The pressure in the chamber was approximately 6$\times10^{-6}$mbar. The applied voltage pulse delivers a 3.5 V square wave with an on time of \SI{200}{ms}. At \SI{3.5}{V} the measured applied current is \SI{3.1}{mA}, resulting in a power dissipation of \SI{10.8}{mW}. The trigger is delayed \SI{50}{ms} from when the PMT starts recording data. The presence of background counts is mainly due to laser scatter inside the chamber. Figure \ref{fig:response} (c) shows that a fluorescence response begins within \SI{10}{ms} after applying a voltage, and after \SI{20}{ms} the fluorescence is already more than halfway towards a steady state count value, which is two orders of magnitude faster than in previous literature\cite{ox}. This can be further improved by applying a higher voltage. The test was repeated with the laser blocked to confirm that the emission detected was from fluorescence, and not faint oven glow. The result was a constant count value across the square wave pulse, further verifying our results. It was also noted that the capping layers had no effect on the fluorescence response and all ovens showed strong fluorescence regardless of the presence of Bi and ITO capping. This further adds to the ability of this oven design to be used with materials such as Ba that oxidise easily in air. 

\section{Conclusion and outlook}

In this paper, we have presented the fabrication of a microfabricated atomic oven with a very fast response time. The oven produces a strong flux of ytterbium at a power dissipation of just \SI{10.8}{mW} as demonstrated with a fluorescence test. The oven is capable of reaching sufficient temperatures to produce a flux for commonly used ions and neutral atoms, such as Ba, Be, Ca, Cd, Mg, Sr, and Yb. Furthermore, elements like barium which oxidise quickly can be covered with a capping layer. While we have not demonstrated this for barium, we demonstrate capping ytterbium ovens with a \SI{\approx5}{nm} layer of ITO and Bi. The capping layer evaporated quickly and did not hinder the observed fluorescence. The microfabricated oven we have demonstrated is designed to be integrated easily into large-scale quantum technologies such as trapped ion quantum computing where spatial constraints, fluorescence response time, and long-term operation capabilities are essential for scaling research into usable technology. We discuss how this device can be easily integrated into a trapped ion microchip with a backside loading slot as small as \SI{15}{\micro\meter} x \SI{5}{\micro\meter} through die bonding and calculate an oven lifetime to produce an atom number density at the ion loading region of \SI{100}{cm^{-3}} to be over ten years of continuous operation.

\section*{Acknowledgments}

Work was carried out at a number of facilities including the Center of MicroNanoTechnology (CMi) at École Polytechnique Fédérale de Lausanne (EPFL) and the London Centre for Nanotechnology (LCN). This work was supported by the U.K. Engineering and Physical Sciences Research Council via the EPSRC Hub in Quantum Computing and Simulation (EP/T001062/1), the U.K. Quantum Technology hub for Networked Quantum Information Technologies (No. EP/M013243/1), the European Commission's Horizon-2020 Flagship on Quantum Technologies Project No. 820314 (MicroQC), the U.S. Army Research Office under Contract No. W911NF-14-2-0106 and Contract No. W911NF-21-1-0240, the Office of Naval Research under Agreement No. N62909-19-1-2116, and the University of Sussex.

\section*{Author declarations}

\subsection*{Conflict of Interest}
The authors have no conflicts to disclose.

\subsection*{Author Contributions}
\textbf{Vijay Kumar} Formal analysis (equal); Investigation (equal); Validation (lead); Visualization (equal); Writing – original draft (equal); Writing – review \& editing (equal).
\textbf{Martin Siegele-Brown} Conceptualization (lead); Investigation (equal); Supervision (lead); Visualization (equal); Writing – original draft (equal); Writing – review \& editing (equal).
\textbf{Parsa Rahimi} Investigation (supporting); Software (lead).
\textbf{Matthew Aylett} Investigation (supporting); Resources (lead).
\textbf{Sebastian Weidt} Conceptualization (equal); Supervision (equal); Writing – review \& editing (equal).
\textbf{Winfried Karl Hensinger} Conceptualization (equal); Funding acquisition (equal); Supervision (equal); Writing – review \& editing (equal).

\section*{Data Availability}

The data that support the findings of this study are available from the corresponding author upon reasonable request. 

\section*{References}
\bibliography{references}

\end{document}